\documentclass[twocolumn,aps,pre,showpacs,amsmath,amssymb,floatfix]{revtex4-1}
\usepackage{graphicx}
\usepackage{dcolumn}
\usepackage{bm,xcolor}

\begin{document}

\title{Echoes from Anharmonic Normal Modes in Model Glasses}

\author{Justin C. Burton}
\email{Author to whom correspondence should be addressed:\newline justin.c.burton@emory.edu}
\affiliation{Department of Physics, Emory University}
\author{Sidney R. Nagel}

\affiliation{James Franck Institute, Enrico Fermi Institute, and Department of Physics, The University of Chicago}

\date{\today}

\begin{abstract} 
Glasses display a wide array of nonlinear acoustic phenomena at temperatures $T\lesssim 1$ K. This behavior has traditionally been explained by an ensemble of weakly-coupled, two-level tunneling states, a theory that is also used to describe the thermodynamic properties of glasses at low temperatures. One of the most striking acoustic signatures in this regime is the existence of phonon echoes, a feature that has been associated with two-level systems with the same formalism as spin echoes in NMR. Here we report the existence of distinctly different type of acoustic echo in classical models of glassy materials. Our simulations consist of finite-ranged, repulsive spheres and also particles with attractive forces using Lennard-Jones interactions. We show that these echoes are due to anharmonic, weakly-coupled vibrational modes, and perhaps provide an alternative explanation for the phonon echoes observed in glasses at low temperatures.
\end{abstract}


\maketitle

\section{Introduction}

Glassy materials behave in a fundamentally different manner than their crystalline counterparts. Perhaps the most striking example is the glass transition; a slow-down of kinetic behavior that strongly depends on temperature \cite{Stillinger2013,Ediger1996}. In contrast to freezing in crystals, there is no sharp phase transition to a solid phase. Rather the time scale for motion of the constituent particles increases smoothly as the temperature is lowered. At low temperatures where glasses are rigid solids, they still retain many properties that are distinct from crystals, yet these properties are seemingly universal among disordered solids. 

%Although there is no well-defined lattice structure in a glass, the thermodynamic properties are still controlled by the density and nature of vibrational modes in the solid. 

It has been known since the early 1970's that the thermodynamic properties of dielectric glasses are different from crystals at temperatures $T\lesssim 1$ K. At these temperatures the heat capacity scales approximately linearly in $T$ and the thermal conductivity scales as $\sim T^2$ \cite{Zeller1971}. For crystals, this scaling is $T^3$ for both quantities. The origin of these differences has traditionally been attributed to a dilute ensemble of two-level tunneling states \cite{Anderson1972,Phillips1972,Phillips1987,Esquinazi1998}. These states are quantum mechanical in nature and spatially localized so that they are weakly coupled to other plane-wave excitations in the solid. Moreover, it has been presupposed that the distribution of two-level energy spacings is very broad, leading to an approximately constant density of levels at low temperatures. Phenomenologically, this picture is consistent with most of the experimental data, yet a fundamental understanding of the origin of these localized modes is still lacking \cite{Leggett2013}. 

Convincing evidence for the existence of two-level tunneling states in glasses comes from acoustic experiments at low temperatures with frequencies $\sim 10^9$ Hz, so that $\hbar\omega\approx k_BT$.  In the experiments an acoustic transducer attached to the glass sample served to both excite the acoustic wave and detect reflections.  Localized modes, as postulated in the quantum-mechanical two-level system model, naturally lead to a rich array of nonlinear acoustic behavior such as a temperature-dependent sound velocity, saturation of attenuation \cite{Hunklinger1972,Arnold1974,Hunklinger1982,Graebner1983,Enss1997,Classen2000,Vural2011}, spectral hole burning and diffusion \cite{Golding1973,Golding1976,Arnold1978,Black1977}.  However, perhaps the most dramatic effect observed in these acoustic experiments was the observation of electric and phonon echoes \cite{Golding1976a,Golding1976,Schickfus1978,Graebner1979,Black1977,Burin2013}.  Since any two-level quantum system has the same dynamics as an isolated spin, these echoes were thought to be analogous to, and have the same formalism as, spin echoes studied in magnetic resonance.  This observation was interpreted as evidence for the quantum mechanical nature of the excitations.  As we will show here, the existence of phonon echoes does not necessitate such an interpretation.  Rather, the echoes can be generated by a distinctly different mechanism that is completely classical in origin and is based on the inherent anharmonicity of the vibrational modes in disordered solids.

In the last decade, there has been a large body of work concerning the vibrational modes in jammed, disordered solids which have helped to shed light on the origins of the thermodynamic properties of glasses \cite{Hecke2010,Liu2010}. An excess density of states at low frequencies naturally arises in jammed systems and relies only on the existence of disorder, not on the details of the particle interaction \cite{OHern2003}. In addition, the well-known ``Boson peak" has been linked to the onset of anomalous modes in jammed systems and other model glasses \cite{Xu2007}. The linear temperature dependence of the thermal conductivity in glasses at intermediate temperatures requires a constant diffusivity; a property that exists in jammed systems above the Boson peak \cite{Xu2009,Vitelli2010}. Finally, at very low frequencies, jammed systems contain quasi-localized, anharmonic vibrational modes which lie at the heart of mechanical rigidity \cite{Xu2010} and indicate the presence of ``soft spots" in amorphous solids \cite{Manning2010,Rottler2014}. 

Unfortunately, inherent difficulties in computing the vibrational mode properties in very large systems have restricted many studies to higher frequencies and temperatures. The low-temperature regime where two-level tunneling states are supposed to dominate the thermodynamic properties has remained elusive. Our goal is not to simulate the largest systems and lowest frequencies directly, but rather to illustrate that the observation of phonon echoes does not require a quantum mechanical interpretation at all. 

The first description of echo phenomena was given be Hahn in 1950 by considering the response of an ensemble of nuclear spins to two excitation pulses separated by a time $\tau$ \cite{Hahn1950,Hahn1953}. Initially, all spins are vertically aligned with an external magnetic field pointing in the $z$-direction. The first pulse acts to rotate the spins towards the $x$-$y$ plane. In between the pulses, the spins precess harmonically at their Larmor frequency, and eventually decohere. The second pulse acts to ``time-reverse'' the system, so that the system becomes coherent again at $t=2\tau$ \cite{Hahn1953}. If we consider an ensemble of spins with different frequencies, the echo manifests as a macroscopic sum of the spin vectors. The maximum echo occurs when the first pulse rotates the spins by  $\theta=\pi/2$ and the second pulse rotates the spins by $\theta=\pi$. A similar mechanism explains photon echoes observed at optical frequencies \cite{Kurnit1964,Abella1966}.

Although less well-known, echo phenomena can also be produced by another mechanism \cite{Gould1965,Kegel1965,Herrmann1969,Korpel1981,Fossheim1982}. One possible mechanism relies on harmonic oscillators that interact in a nonlinear way with the excitation pulses, as in temperature quench echoes \cite{Nagel1983,Xu1995}. Another possible mechanism involves anharmonic oscillators whose resonant frequency shifts with increasing amplitude. This mechanism is the source of observed echoes in many different systems ranging from cyclotron modes in plasmas \cite{Hill1965,Gould1965,Kegel1965,Herrmann1967} to the vibrations of individual particles in piezoelectric powders \cite{Kajimura1976,Fossheim1978,Kuindersma1976}. Although there are many similarities between anharmonic echoes and spin echoes, there are many characteristic differences such as the relationship between the echo amplitude and pulse spacing, $\tau$, and the existence of multiple echoes after only two pulses for anharmonic echoes. A comprehensive review on both types of echoes can be found in reference \cite{Fossheim1982}.

%With anharmonic echoes, the details of the individual excitation pulses are less important, and it is the time evolution between the pulses which results in the echo.

\begin{figure}
\begin{center}
\includegraphics[width=.48 \textwidth]{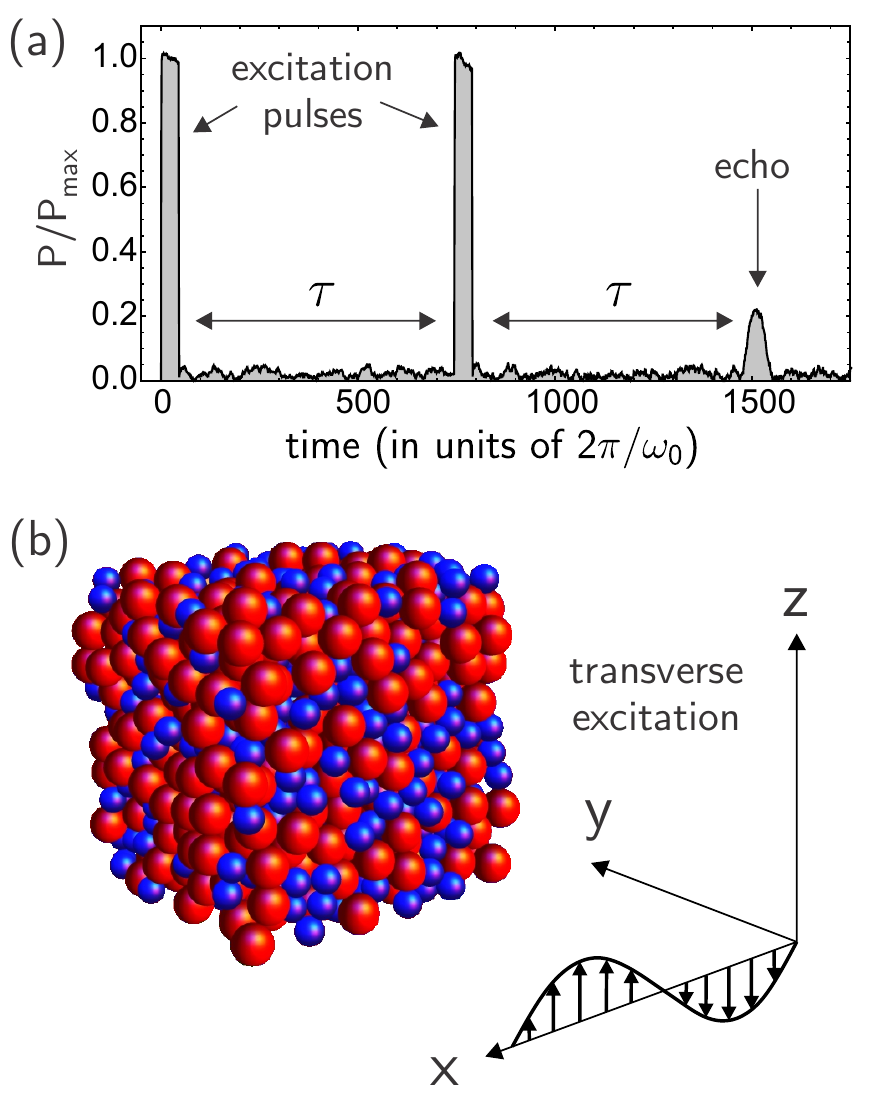}
\caption[]{(Color online) (a) Acoustic echo generated in jammed systems composed of 1000 particles at $\phi$ = 0.70. The result is an average over the response of 10000 independent systems. After a series of excitation pulses separated by time $\tau$, a spontaneous re-phasing of the vibrational modes occurs at a time $t=2\tau$. (b) Visual representation of one of the jammed systems used in generating the echo. The red particles are 1.4 times larger than the blue particles. The acoustic pulses are excited by a transverse standing wave along the $x$-direction.
} 
\label{setup}
\end{center} 
\end{figure}

In this paper we use simulations of model glasses to show that classical vibrational modes in disordered solids can act as weakly-coupled anharmonic oscillators, and when excited by a series of pulses, produce echoes similar to those seen in experiments in glasses at low temperatures (Fig.\ \ref{setup}a). By varying the pulse amplitude, spacing, and number of pulses, we can compare our results directly to experimental data. Our simulations are performed with both finite-ranged repulsive spheres and particles with Lennard-Jones interactions. 

%The results agree with and support the emerging picture from studies of jammed systems: that low-frequency, quasi-localized modes can serve as an origin of weakly-coupled ``resonant modes'' described in the glass literature, and that the inherent anharmonicity of these modes can give rise to nonlinear acoustic properties. 

\section{Numerical Model}
\label{NumericalModel}

The majority of our simulations consist of a 3-dimensional ensemble of frictionless, spherical particles with finite-range, repulsive interactions \cite{OHern2003}. We use a 50-50 binary mixture of particles with two radii, $\sigma$ and 1.4$\sigma$. All particles have a mass $m$. The pair-potential between any two particles is given by the following:

\begin{equation}
V(r_{ij})= 
\begin{cases} 
\dfrac{2\epsilon}{5}\left(1-\dfrac{r_{ij}}{\sigma_i+\sigma_j}\right)^{5/2}
 & \text{$r_{ij}<\sigma_{i}+\sigma_{j}$,}
\\
0 &\text{$r_{ij}\geq\sigma_{i}+\sigma_{j}$,}
\end{cases}
\end{equation}
where $r_{ij}$ is the distance between the centers of particles $i$ and $j$, and $\epsilon$ is the energy scale of the interaction. All quantities reported here have lengths measured in units of $\sigma$, mass in units of $m$, and frequency in units of $\sqrt{\epsilon/m\sigma^2}$. The 5/2-exponent in the potential is derived from linear, elastic, Hertzian contact mechanics of spherical particles. An important feature of this type of potential is the natural nonlinearity of inter-particle interaction. There will always be a nonlinear correction to the harmonic approximation for the potential energy, so that the frequency response of individual modes will vary with vibrational amplitude. This feature will turn out to be essential for the generation of anharmonic echoes, and will be discussed in section \ref{mode_char}. 

Individual systems were created by randomly placing $N$ particles in a cubic box with periodic boundary conditions on all sides, which represents a dense gas at $T=\infty$. Each system is then quenched to $T=0$ at the nearest local potential-energy minimum using the Fast Inertial Relaxation Engine (FIRE) algorithm \cite{Bitzek2006}. The resulting state of the system (i.e., jammed or un-jammed) will depend on the volume fraction $\phi$ of particles. For the size ratio and particle interactions studied here, jamming occurs at $\phi= \phi_c\approx0.64$ \cite{OHern2003}. All simulations reported here are for a volume fraction $\phi = 0.70$, so that the systems are well into the jammed regime. In addition, at this volume fraction, approximately 0.3\% of the particles have no overlaps after the initial quench. These particles contribute trivial zero-frequency modes to the system, so they are removed prior to acoustic excitation. A 3-D representation of a quenched 1000-particle system is shown in Figure \ref{setup}b. 

In addition to finite-ranged, repulsive interactions, we also simulated Kob-Andersen binary Lennard-Jones systems \cite{Kob1995}.  Each system consists of 800 $A$ and 200 $B$ particles with equal mass $m$ interacting in three dimensions.  The pair-potential between particles is given by:
\begin{equation}
V(r_{ij})=\dfrac{\epsilon_{ij}}{72}\left[\left(\dfrac{\sigma_{ij}}{r_{ij}}\right)^{12}-\left(\dfrac{\sigma_{ij}}{r_{ij}}\right)^{6}\right],
\label{pot}
\end{equation}
where $\epsilon_{AB}=1.5\epsilon_{AA}$, $\epsilon_{BB}=0.5\epsilon_{AA}$, $\sigma_{AB}=0.8\sigma_{AA}$, and $\sigma_{BB}=0.88\sigma_{AA}$ \cite{Xu2007}. The potential is cut off at $r_{ij}=2.5\sigma_{ij}$ and the potential is shifted so that $V(2.5r_{ij})=0$. We also add an additional linear correction so that $V'(2.5r_{ij})=0$. All Lennard-Jones systems were created at a density $\rho = 1.2$, then quenched using the FIRE algorithm.

Once a system is quenched, we excited the vibrational modes using external pulses. Each pulse consisted of applying a transverse, spatially-varying sinusoidal force to the particles. The force on the $i$th particle is given by:
\begin{equation}
\vec{\bf F}_i=F_0\sin(k_x x_i-\omega_0 t)\hat{\bf y},
\label{force}
\end{equation}
where $k_x=2\pi/L$ is the wavevector, and $L$ is the box size. We used the longest wavelength that could fit along one boundary of the domain (as shown in Figure \ref{setup}b). In order to maximize the coupling of the pulse to a narrow band of vibrational modes, the pulse frequency was restricted so that $\omega_0/k_x\approx v_s$, where $v_s$ is the speed of sound in the system. The pulse amplitude $F_0$, and duration $t_p$ were adjustable parameters, although typical ranges of $t_p$ were 10-45 cycles, where the period of 1 cycle = $2 \pi/\omega_0$. Longitudinal polarizations were also studied with qualitatively similar results, yet the transverse excitations were better coupled to the anharmonic, low-frequency modes in the systems. Thus the majority of simulations used external pulses according to Eq.\ \ref{force}. 

The response of the system can be measured in many different ways. We chose perhaps the most natural way, and measured the response along the same vector that defined the excitation. That is, the forcing $\vec{\bf F}$ represents a vector with 3$N$ elements, and can be expanded in eigenmodes of the system. If the modes do not couple, then the total energy in each mode remains constant in time. The most convenient way to access the response of the excited modes was to measure the power $P=\vec{\bf v}\cdot\vec{\bf F}$, where $\vec{\bf v}$ is the velocity vector of the particles. The resulting signal was then normalized by the maximum power ($P_{max}$) during the pulses, as shown in Figure \ \ref{setup}a.

For systems quenched from initially random positions ($T=\infty$), we found that some modes often went unstable during excitation by an acoustic pulse. This is likely due to the crossing of a significant energy barrier in the system, and was followed by a $\approx10-20\%$ drop in the potential energy of the system. Upon re-quenching the system following such an instability, the minimum potential energy at $T=0$ also decreased by $\approx10-20\%$. The excess potential energy is likely due to the preparation of the system by quenching from $T=\infty$, without any annealing steps. By repeatedly pulsing each system with acoustic pulses of decreasing $F_0$, with each pulse followed by a quench, we found that the stability of the system increased dramatically. Thus all data reported here comes from systems which have been prepared using this annealing protocol.

\section{Characterization of Modes in Jammed Systems}
\label{mode_char}

\begin{figure}
\begin{center}
\includegraphics[width=.48 \textwidth]{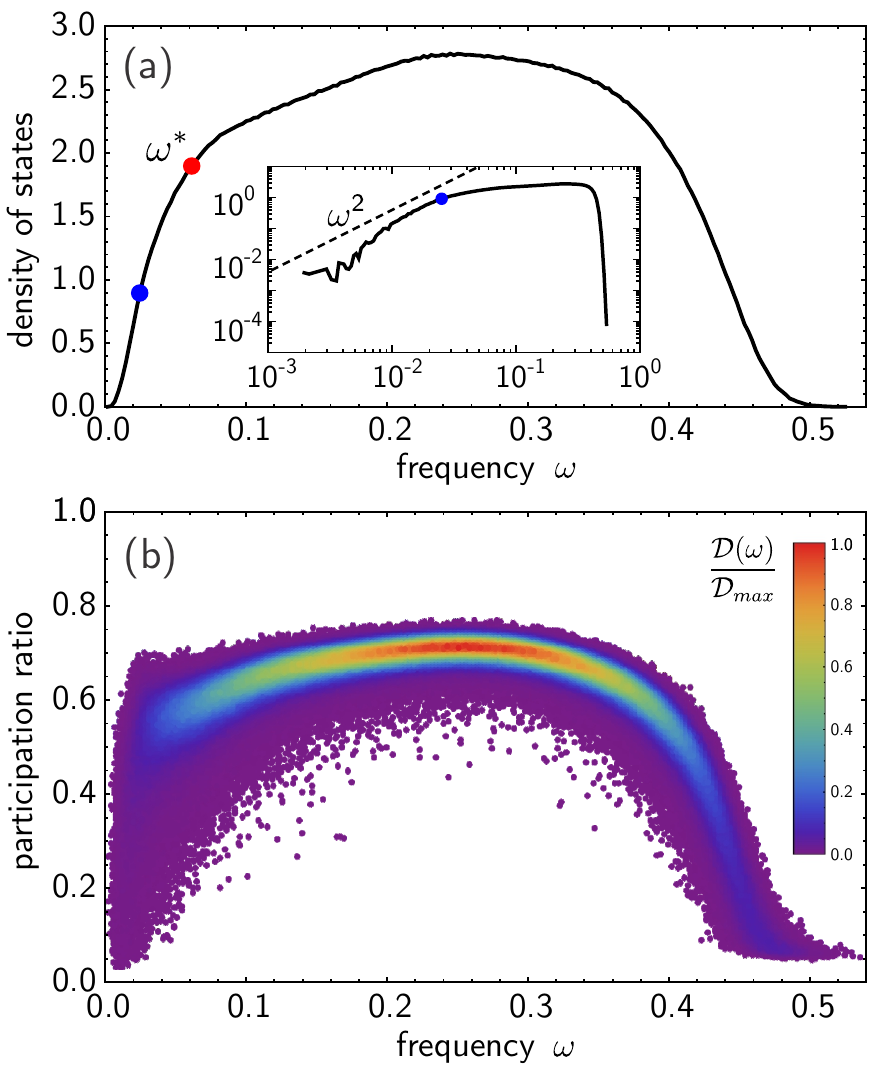}
\caption[]{(Color online) (a) Density of states for systems of 1000 bi-disperse particles at $\phi$=0.70. The solid black lines are the average of 1000 independent systems. The inset shows the same data on a log scale. The dashed line represents an $\omega^2$ behavior, consistent with Debye theory. The red dot shows the approximate position of $\omega^{*}$, and the blue dot shows the frequency at which most of the simulations are performed. (b) Participation ratio of all modes in 1000 independent systems. The color scale represents the density of the points. Red is the maximum density of states $\mathcal{D}_{max}$, and purple is near zero. } 
\label{DOS}
\end{center} 
\end{figure}

One of the defining characteristics of crystalline elastic solids is that at sufficiently low frequencies, all vibrational modes are plane-wave acoustic modes. In stark contrast, glasses display an excess of anomalous modes at low frequencies, some of which are spatially localized. Jammed systems of soft, frictionless spheres contain a very large number of low-frequency modes as well \cite{Liu2010}. The amount of excess modes will depend on the distance from the critical volume fraction,  $\Delta\phi=\phi-\phi_c$, where $\phi_c$ is the volume fraction when the system first begins to jam. The peak in the excess number of modes (which is known as the Boson peak in the glass literature) occurs at a characteristic frequency $\omega^*$, which tends towards zero as $\phi_c$ is approached: $\omega^*\propto\Delta\phi^{3/4}$ for the potential chosen in equation \ref{pot} \cite{Liu2010}. 

Figure \ref{DOS}a shows the average density of states for systems of $N$ = 1000 particles. The approximate location of $\omega^*$ is shown by the red point. The inset shows the same data on a log scale. The modes below $\omega^*$ consist of a mix of extended and quasi-localized vibrational modes \cite{Silbert2009}. Since each system only contains 1000 particles, the lowest frequency plane-wave mode would occur at $\omega\approx0.025$. 

The degree of localization of the modes is illustrated in Figure \ref{DOS}b, which shows the participation ratio for each mode. The participation ratio measures the fraction of particles participating in a given vibrational mode:
\begin{align}
\label{prat}
p(\omega_m)=\frac{\left(\sum_l\left|\hat{\bf e}_{m,l}\right|^2\right)^2}{N\sum_l\left|\hat{\bf e}_{m,l}\right|^4},
\end{align}
where $\hat{e}_{m,l}$ is the $l$th component of the unit eigenvector corresponding to the $m$th eigenmode. At low frequencies, the jammed systems contain a broad distribution of participation ratios, as shown by the large spread in purple data points. Many of these modes are quite localized (low participation ratio). It has also been shown that these modes exist independent of $\Delta\phi$ \cite{Xu2010}.

One natural consequence of spatial localization in a vibrational mode is that for a given amount of energy, fewer particles are undergoing a larger amplitude motion. This larger amplitude induces nonlinear effects in the vibration at smaller energies. Low-frequency plane waves in crystalline solids are spatially extended and are the most harmonic modes in the system due to the small relative displacements between neighboring particles. However, in jammed solids, the lowest frequency modes are the most {\it anharmonic} \cite{Xu2010}. 

\begin{figure}
\begin{center}
\includegraphics[width=.48 \textwidth]{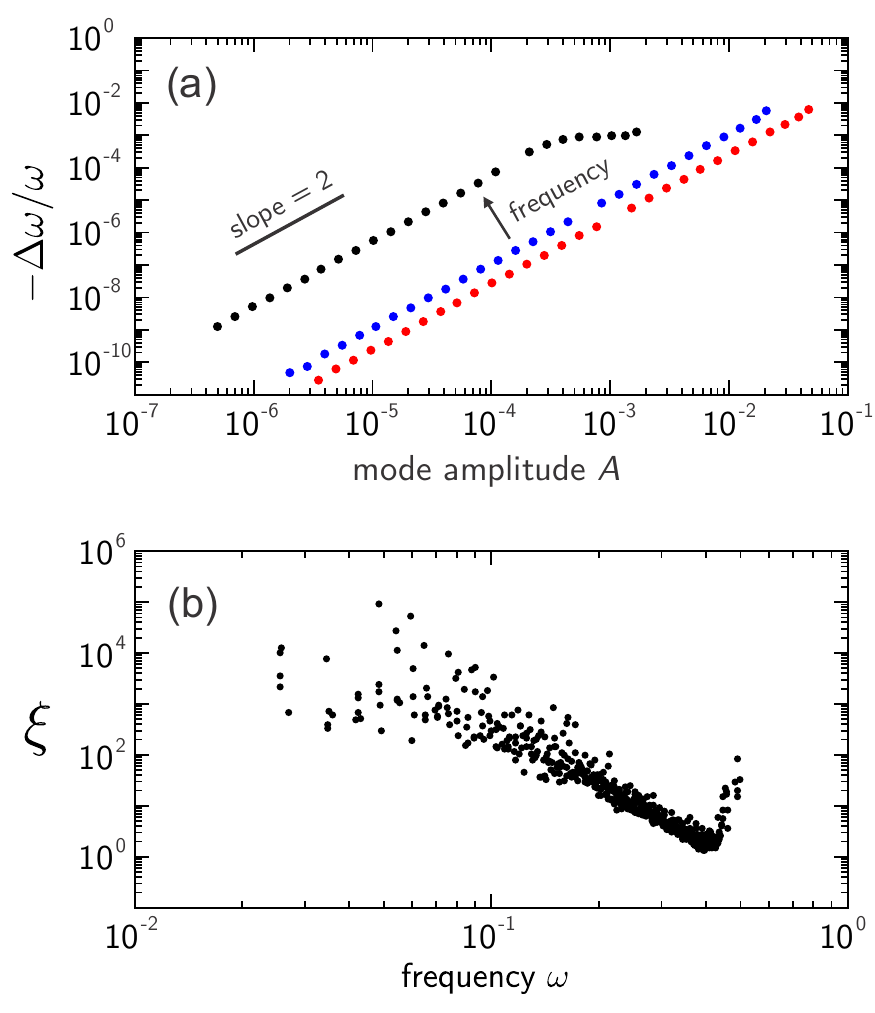}
\caption[]{(Color online) (a) Normalized frequency shift versus amplitude of 3 modes in a single 1000-particle system. Over a broad range of amplitudes, equation \ref{freqshift} is valid at high (red), intermediate (blue), and low (black) frequencies. Deviations at higher amplitudes are due to coupling between nearby modes. (b) Coefficient of frequency shift $\xi$ versus mode frequency for 500 modes in 5 different 1000-particle systems. For a given amplitude $A$, modes at lower frequency will experience a larger frequency shift.}
\label{anharmonicity}
\end{center} 
\end{figure}

\subsection{Anharmonic Frequency Shifts}

An important consequence of nonlinearity is that the fundamental frequency of the mode will shift with amplitude. Let us consider the simplest model of an anharmonic oscillator with mass $m$ and fundamental frequency $\omega$ with a cubic perturbation to the potential $V$:
\begin{equation}
V=\dfrac{m\omega^2}{2}x^2+\dfrac{m\omega^2}{3x_0} x^3.
\end{equation}
The equation of motion of this oscillator is thus
\begin{equation}
m\ddot{x}=-m\omega^2x\left(1-\dfrac{x}{x_0}\right).
\end{equation}
To second order, it can be shown that the frequency of the oscillator depends on the square of the amplitude of vibration \cite{Goldstein1980,Apostol2005}:
\begin{equation}
\dfrac{\Delta\omega}{\omega}=-\xi A^2,
\label{freqshift}
\end{equation}
where $\xi=5/12x_0^2$. 

We can measure the anharmonicity of the modes in jammed systems by applying an initial amplitude to the modes at $t=0$ with all particles at rest, then letting the system evolve in time. Specifically, this is accomplished by adding a vector $A\hat{\bf e}_m$ to the initial position vector of all of the particles, where $A$ is the amplitude and $\hat{\bf e}_m$ is the eigenvector associated with the $m$th mode. After 200 cycles, the resulting motion of the particles along $\hat{\bf e}_m$ is fit to a sinusoidal function to obtain the frequency of the mode.

Figure \ref{anharmonicity}a shows the normalized frequency shift as a function of amplitude for three modes at high, intermediate, and low frequencies. The frequency shift is quadratic in amplitude, and is larger at low frequencies. At higher amplitudes, the interaction among nearby modes becomes pronounced and energy is transferred between modes, leading to a damping of the vibrations and other forms of nonlinearities. Eventually, at very large amplitudes, particle rearrangements occur and the eigenmodes have changed, so our analysis is no longer valid.

Figure \ref{anharmonicity}b shows how $\xi$ depends on mode frequency. Localized modes at high are anharmonic, but the most anharmonic modes lie at low frequencies where the modes are quasi-localized. The broad distribution of $\xi$ at lower frequencies is related to the broad distribution in the participation ratio. 

\section{Echoes from Anharmonicity}

In order to understand how an ensemble of anharmonic oscillators can give rise to an echo, let us first consider the response of a single oscillator to two excitation pulses. Our derivation is similar to previous derivations \cite{Kegel1965,Herrmann1967,Fossheim1978}, except that here we explicitly deal with mechanical oscillators for arbitrary amplitudes. We will then sum the contribution of many single oscillators with different natural frequencies. For simplicity, we will only consider $\delta$-function pulses which add a finite amount of energy to the oscillator in a short period of time. 

At time $t=0$, the first pulse excites the oscillator so that it begins with amplitude $A_1$, then evolves in time:
\begin{equation}
x(0<t<\tau)=A_1e^{i\omega t(1-\xi A_1^2)}.
\end{equation}
Here the frequency is slightly less than the fundamental frequency due to the finite amplitude, so that $\omega_1=\omega(1-\xi A_1^2)$. We have also ignored any other higher harmonics in the solution stemming from the nonlinearity of the oscillator, and only consider the frequency shift to the fundamental mode. 

At time $t=\tau$, we apply a second pulse, which adds amplitude $A_2$ to the position of the oscillator:
\begin{equation}
x(\tau)=A_2+A_1e^{i\omega \tau(1-\xi A_1^2)}. 
\end{equation}
For simplicity, we will assume that $A_1$ and $A_2$ are real, although the same analysis can be done in the case that they are complex. The new amplitude of oscillation is:
\begin{equation}
|x(\tau)|^2=A_1^2+A_2^2+2A_1A_2\cos\left(\omega t\left(1-\xi A_1^2\right) \right).
\label{secondamp}
\end{equation}
The evolution of the oscillator after the second pulse depends on its amplitude, so that
\begin{align}
\label{fullsol}
&x(t\geq\tau)=\\
&\left(A_2+A_1e^{i\omega \tau\left(1-\xi A_1^2\right)}\right)e^{i\omega (t-\tau)\left(1-\xi |x(\tau)|^2\right)}\nonumber.
\end{align} 
Combining equation \ref{secondamp} and \ref{fullsol}, we obtain the full solution of the oscillator when $t\geq\tau$:
\begin{align}
\label{newsol}
&x(t\geq\tau)=\\
&\left(\left(A_2+A_1e^{i\omega \tau\left(1-\xi A_1^2\right)}\right)e^{i\omega (t-\tau)\left(1-\xi\left(A_1^2+A_2^2\right)\right)}\right)\times\nonumber\\
&e^{-i2\xi\omega(t-\tau)A_1A_2\cos\left(\omega\tau\left(1-\xi A_1^2\right)\right)}.\nonumber
\end{align}

The complexity here is due to the fact that there is a cosine function in the argument of the exponential. We can simplify this part by use of the Jacobi-Anger expansion \cite{Abramowitz1964}:
\begin{align}
e^{iz\cos\theta}=\sum_{n=-\infty}^{\infty}i^nJ_n(z)e^{in\theta},
\end{align}
where $J_n$ is a Bessel function of the 1st kind. Then equation \ref{newsol} becomes:
\begin{align}
\label{finsol}
&x(t\geq\tau)=\\
&\left(\left(A_2+A_1e^{i\omega \tau\left(1-\xi A_1^2\right)}\right)e^{i\omega (t-\tau)\left(1-\xi\left(A_1^2+A_2^2\right)\right)}\right)\times\nonumber\\
&\sum_{n=-\infty}^{\infty}i^nJ_n\left(2\xi\omega(\tau-t)A_1A_2\right)e^{-in\omega\tau\xi A_1^2}e^{in\omega\tau}\nonumber.
\end{align}
At this point it is helpful to define a characteristic frequency shift $\Omega=\omega \xi A_1^2$ and pulse amplitude ratio $\alpha=A_2/A_1$. With these substitutions and some algebraic manipulations, equation \ref{finsol} becomes:
\begin{align}
\label{finalsol}
&x(t\geq\tau)=\\
&\sum_{n=-\infty}^{\infty}e^{i\omega(t-n\tau)}i^{3n}A_1e^{i\Omega(n\tau-t)}e^{i\Omega\alpha^2(\tau-t)}\times\nonumber\\
&\left[J_n(2\alpha\Omega(t-\tau))+i\alpha J_{n-1}(2\alpha\Omega(t-\tau))\right]\nonumber.
\end{align}

We may now identify terms in the solution that vary on different time scales. Oscillatory terms containing ``$\omega t$" in their argument vary rapidly in time, whereas terms with ``$\Omega t$" will vary much more slowly since the frequency shift is much smaller than the fundamental frequency ($\Omega/\omega\ll1$). Thus we can write
\begin{align}
\label{fastslow}
x(t\geq\tau)=\sum_{n=-\infty}^{\infty}G\left(\Omega t\right)e^{i\omega(t-n\tau)}
\end{align}
where 
\begin{align}
\label{fullG}
&G(\Omega t)=i^{3n}A_1e^{i\Omega(n\tau-t)}e^{i\Omega\alpha^2(\tau-t)}\times\\
&\left[J_n(2\alpha\Omega(t-\tau))+i\alpha J_{n-1}(2\alpha\Omega(t-\tau))\right]\nonumber.
\end{align}

Equations \ref{fastslow} and \ref{fullG} apply to only a single oscillator which is excited by two delta function pulses. An echo involves the coherent sum of many oscillators at a given point in time. Each oscillator may have a different fundamental frequency, $\omega$. Thus the echo amplitude, $X$, will be given by
\begin{align}
\label{fullecho}
X(t\geq\tau)=\sum_{m}\sum_{n=-\infty}^{\infty}G\left(\Omega_m t\right)e^{i\omega_{m}(t-n\tau)},
\end{align}
where $\omega_{m}$ is the fundamental frequency of the $m$th oscillator, and $\Omega_m=\omega_{m}\xi A_1^2$.
When performing the sum over $m$, the exponential term $e^{i\omega_{m}(t-n\tau)}$ will vary rapidly with time and sum to zero since $\omega$ is different for every oscillator, i.e., the total signal will be decoherent. However, if $t=n\tau$, the exponential term will be near unity and the oscillators will be coherent. The echo amplitude, $X$ will then depend only on $G(\Omega_m t)$, which varies slowly with time since $\Omega_m\ll\omega_{m}$. One important consequence of equation \ref{fullecho} is that not only do we expect an echo at $t=2\tau$, but also multiple echoes at $t=3\tau,$ $4\tau,$ etc. This is a distinguishing feature of classical echoes in anharmonic oscillators. The simplest description of a quantum mechanical spin echo only contains features at $t=2\tau$. At later times the precessing spins become incoherent. 

\begin{figure}
\begin{center}
\includegraphics[width=.48 \textwidth]{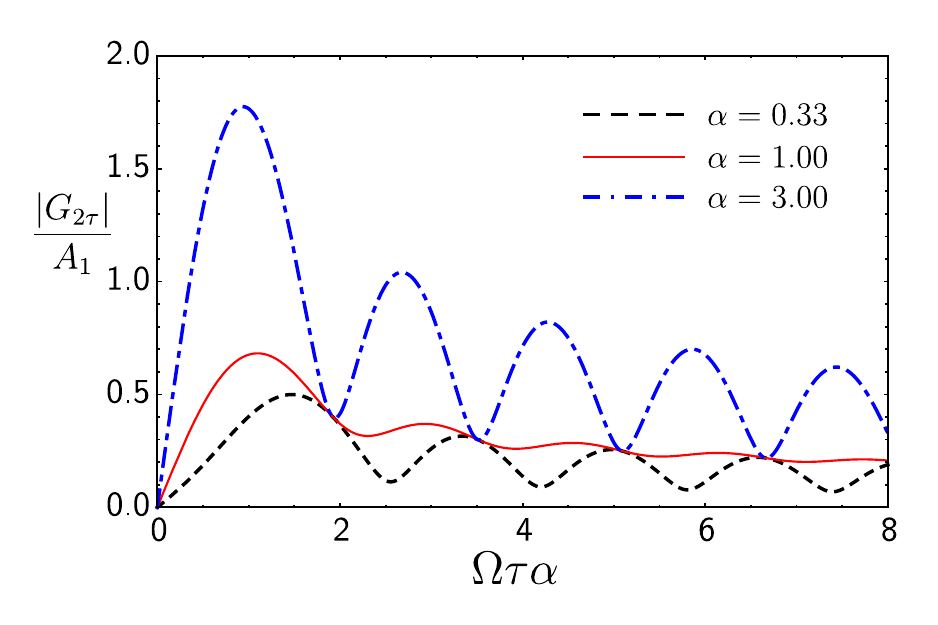}
\caption[]{(Color online) Proxy for echo amplitude $|G_{2\tau}|$, normalized by the first pulse amplitude $A_1$, as a function of $\Omega\tau\alpha$ (see equation \ref{echoamp}). Over a broad range of values for the ratio of pulse amplitudes $\alpha$, the maximum echo occurs for $1<\Omega\tau\alpha<2$. }
\label{analytic}
\end{center} 
\end{figure}

Let us assume that the excitation pulse excites a narrow band of oscillators with similar anharmonicity, so that $\Omega_m$ is approximately constant, and we may drop the subscript $m$. If we only consider the first echo, i.e. $t=2\tau$ and $n=2$, then $G$ becomes:
\begin{align}
\label{echoamp}
G_{2\tau}=-A_1e^{-i\Omega\tau\alpha^2}\left(i\alpha J_{1}(2\Omega\tau\alpha)+J_{2}(2\Omega\tau\alpha)\right).
\end{align}
so that the echo amplitude is approximately given by
\begin{align}
\label{absechoamp}
X\propto|G_{2\tau}|=A_1\sqrt{\alpha^2 J_{1}(2\Omega\tau\alpha)^2+J_{2}(2\Omega\tau\alpha)^2}.
\end{align}
In this form it is apparent that the echo amplitude depends on the pulse spacing, in contrast to spin echoes. Specifically, the echo amplitude tends to zero at small pulse spacings. This can be seen by considering the frequency shift of the oscillators as a slowly-varying phase. If the phase does not have time to evolve between the excitation pulses, then its effect on the dynamics will be reduced. The appearance of multiple echoes and the dependence on pulse spacing will be discussed in more detail in section \ref{results}.

Figure \ref{analytic} shows shows how equation \ref{echoamp} depends on the quantity $\Omega\tau\alpha$. Of particular importance is where the maximum echo is located. When the second pulse amplitude is comparable to the first pulse amplitude (i.e $\alpha\approx 1$), then the maximum echo is achieved when $\Omega\tau\approx 1$. This means that the characteristic frequency shift of the oscillators should be the inverse of the pulse spacing. The requirements to achieve the maximum echo amplitude are technically different in spin echoes, where the second pulse ($\pi$-pulse) should be twice as large as the first ($\pi/2$-pulse), given they are the same duration.  

%Finally, we can compare our result to previous authors  we assume a very weak anharmonicity such that $\Omega\tau\alpha\ll1$, then we obtain the result
%\begin{align}
%G_{2\tau}=-iA_1\Omega\tau\alpha^2=-iA_1A_2^2\xi\omega_0\tau,
%\end{align}
%which is identical to the expression derived in reference \cite{Herrmann1969}.

\section{Echoes in Model Glasses}
\label{results}

The discussion up to this point has only considered isolated, independent, anharmonic oscillators. We now turn our attention to echoes in model glasses. We emphasize here that each ``oscillator'' is a normal vibrational mode of the disordered solid. The echo signal is the sum of the vibrational motion of all of the excited modes. For very small amplitudes, each mode is linearly independent. However, for larger amplitudes they will necessarily couple energy between different modes, invalidating our analysis in the previous section. As illustrated by Figure \ref{anharmonicity}, the normal modes are naturally anharmonic, so that an echo should be observable so long as the amplitude of each mode is not too large, and they remain linearly independent.

When excited by an acoustic pulse near $T=0$, a single vibrational mode will increase in amplitude, and the final amplitude of vibration will depend on difference between the frequency of the oscillator and the frequency of the exciting pulse, in addition to the spatial coupling to the polarization of the excitation (equation \ref{force}). The number of modes excited by a given pulse is inversely proportional to the duration of the pulse. For long pulses, only modes with frequencies near the excitation frequency will be driven to large amplitudes, whereas for short pulses, many modes of different frequencies will be excited (e.g., a delta-function pulse will excite modes of all frequencies equally). 

\begin{figure}
\begin{center}
\includegraphics[width=.48 \textwidth]{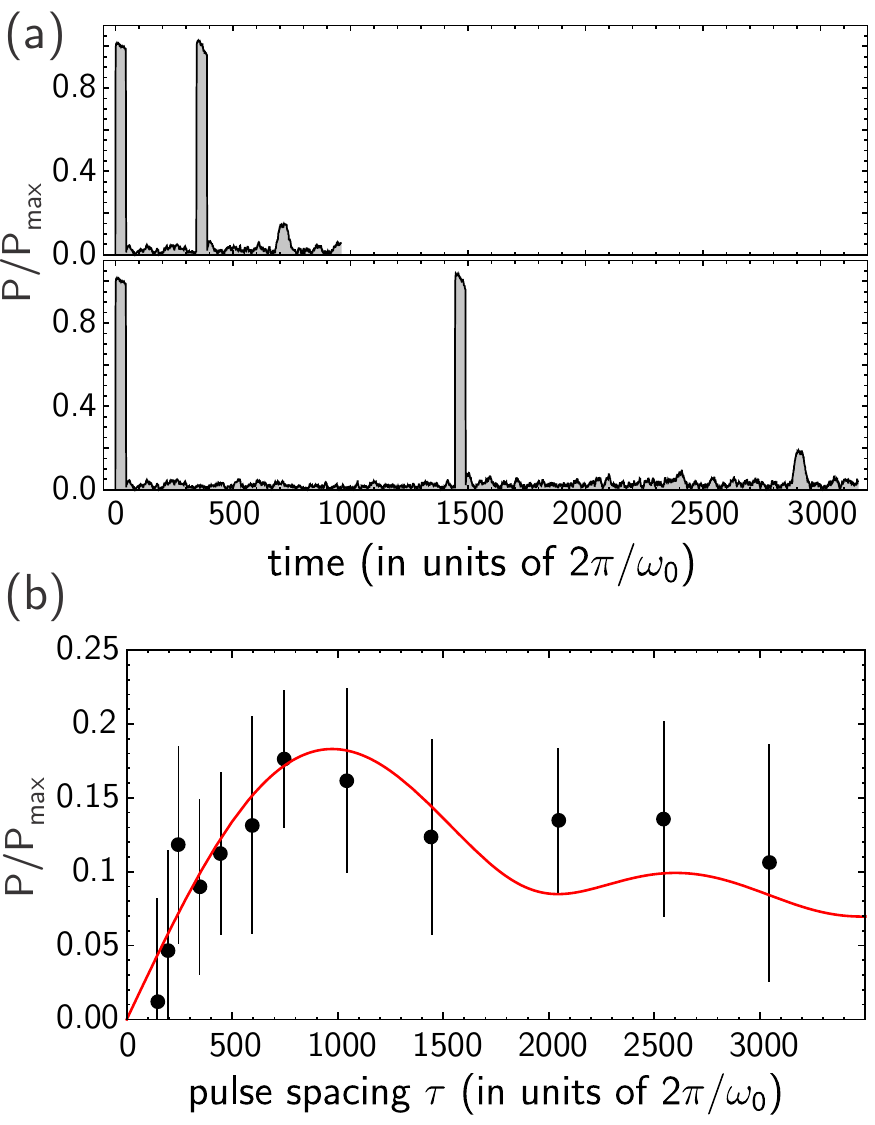}
\caption[]{(Color online) (a) Echo amplitude, normalized by $P_{max}$, vs. time for two different values of pulse spacing $\tau$. A third example, with an intermediate value of $\tau$, is shown in Figure \ref{setup}a. Each signal is the average of 10,000 independent systems, each composed of 1000 particles. The pulse width is $t_p$ = 45 cycles, and the pulse frequency is $\omega_0$ = 0.025. (b) Normalized echo amplitude vs. $\tau$. The error bars represent the size of the noise between the second pulse and the echo. The red line is a fit to the data using equation \ref{fitform}.} 
\label{echovstau}
\end{center} 
\end{figure}

A second pulse at a later time can either increase or decrease the amplitude of an individual mode, depending on the phase difference between the mode and the excitation. An echo will be formed by the average of an ensemble of vibrational modes, which become coherent at a later point in time. For systems composed of 1000 particles, we found that averaging over 10,000 independent systems was necessary in order to achieve a sufficient echo signal above the background noise. Figure \ref{echovstau}a shows the averaged amplitude at two different values of pulse separation $\tau$. Both pulses have identical amplitudes ($F_0\approx 5\times10^{-5}$), and identical pulse durations: $t_p=45$ cycles (this value for $t_p$ was chosen because it was close to the value used in the original experiments which observed phonon echoes at low temperatures in glasses \cite{Graebner1979}). In both plots, the echo is apparent at $t\approx 2\tau$. Taking into account the finite pulse width ($t_p$), the exact position of the echo is $2\tau+2t_p$, since $\tau$ is measured from the center of each excitation pulse and the first pulse begins at $t=0$. 

Figure \ref{echovstau}b shows the echo amplitude (normalized by $P_{max}$) as $\tau$ is varied. This dependence can be understood using equation \ref{absechoamp}. We fit the data to the form:
\begin{align}
\label{fitform}
|A_{2\tau}|=K_1\sqrt{J_{1}(K_2\tau)^2+J_{2}(K_2\tau)^2},
\end{align}
 where $K_1$ and $K_2$ are fitting parameters. The best fit is shown by the red line in Figure \ref{echovstau}. This is essentially the same curve as the red line in Figure \ref{analytic}. The error bars represent the average noise in the amplitude in the region between the second pulse and the echo. Although equation \ref{fitform} is derived from the dynamics of a single oscillator, both $K_1$ and $K_2$ represent an average over the different modes excited by the pulses. Since $K_1\propto A_1$, and $A_1$ represents the initial amplitude, its value will vary considerably from mode to mode. However, $K_2$  will be more uniform since it represents the frequency shift, $\Omega$, and only modes that satisfy $\Omega\tau\approx 1$ will contribute to the echo.

\begin{figure}
\begin{center}
\includegraphics[width=.48 \textwidth]{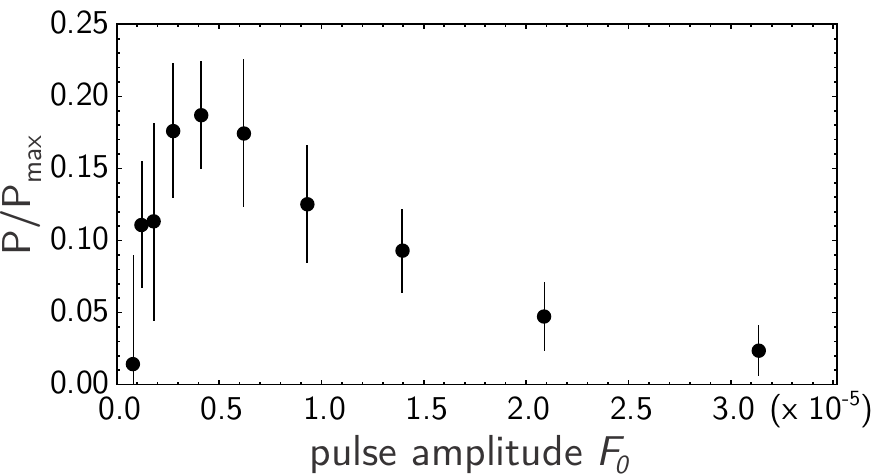}
\caption[]{Echo amplitude versus pulse amplitude $F_0$. Each data point is the average of 10,000 independent systems, each composed of 1000 particles. The pulse width is $t_p$ = 45 cycles, and the pulse frequency is $\omega_0$ = 0.025, which is below $\omega^*$ as shown in Figure \ref{DOS}a. The decay at long times is due to nonlinear coupling between the modes which causes energy to spread to other modes in the system.} 
\label{echovsamp}
\end{center} 
\end{figure}

The reasonable agreement in Figure \ref{analytic} is due to the fact that the condition $\Omega\tau\approx 1$ can be achieved by increasing $\tau$ rather than the amplitude, so that the nonlinearity remains a perturbation to the system and their is little cross-talk between adjacent modes. However, if we vary the pulse amplitude instead of $\tau$, then we inject more energy into each mode at higher amplitudes and the echo amplitude is reduced due to nonlinear couplings between modes, which is not accounted for in the model. Figure \ref{echovsamp} shows a peak in the echo amplitude as the pulse amplitude is varied. This is expected from equation \ref{absechoamp}, and illustrated in Figure \ref{analytic}. However, the data in Figure \ref{echovsamp} decays much more rapidly, which is likely due to the coupling between modes for larger pulse amplitudes, where energy is being redistributed to other modes in the system.

\begin{figure}
\begin{center}
\includegraphics[width=.48 \textwidth]{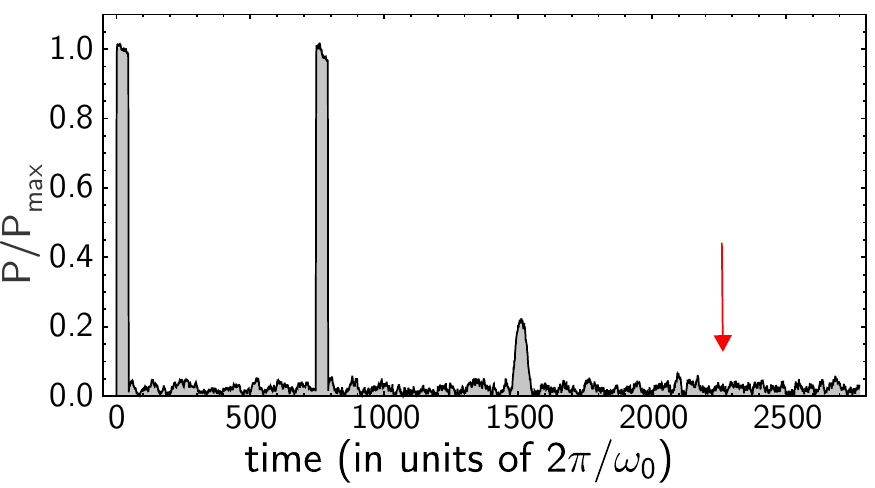}
\caption[]{Multiple echoes are not observable in the average response of 10,000 systems due to noise limitations. Each system has 1000 particles. The pulse amplitude and frequency are the same as in Figure \ref{setup}a and Figure \ref{echovstau}a. The red arrow indicates the predicted position of the echo at $t=3\tau$.} 
\label{multiple}
\end{center} 
\end{figure}

When compared to spin echoes, a defining characteristic of anharmonic echoes is the occurrence of multiple echoes after just two pulses (equation \ref{fastslow}). Figure \ref{multiple} shows the average response of 10,000 systems of 1000 particles each, identical to Figure \ref{setup}a and Figure \ref{echovstau}a, except extended to longer times. Multiple echoes are clearly not visible. This is likely due to the signal noise in this region. One possible remedy is to average over many more systems, since the noise decreases as $\sqrt{N}$, although this was computationally prohibitive. One may reasonably expect the $3\tau$ echo to be reduced in amplitude by the same factor as the $2\tau$ echo is with respect to the pulse amplitude. If this is true, then it is not surprising that the $3\tau$ echo is not visible since it would clearly lie below the noise.

\begin{figure}
\begin{center}
\includegraphics[width=.48 \textwidth]{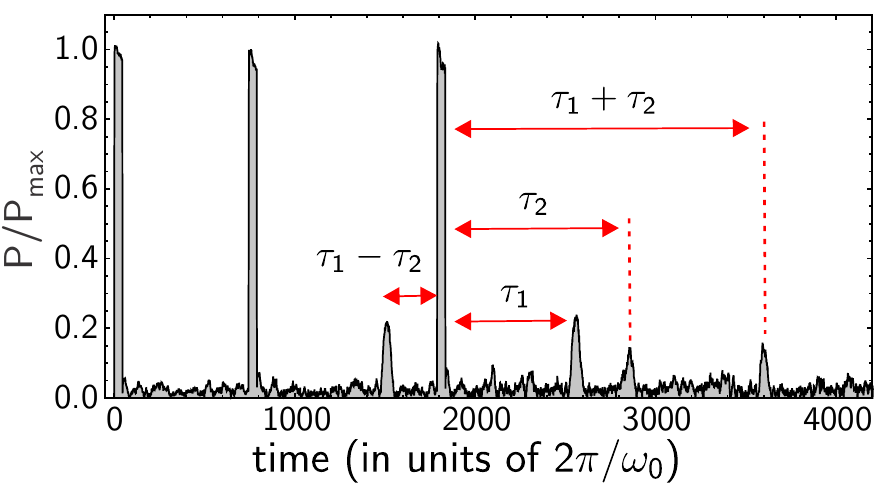}
\caption[]{(Color online) Echo signal from 10,000 averaged systems, as in Figure \ref{multiple}, with the addition of a third pulse after the first echo. The three-pulse echo sequence produces four total echoes, with the positions indicated by the red arrows. This is a characteristic of both parametric (spin) echoes and anharmonic echoes.} 
\label{3pulse}
\end{center} 
\end{figure}

If one applies a third pulse to the system, then there will be a total of four echoes that can be observed. The positions of these echoes are $\tau_1$, $\tau_2$, $\tau_1+\tau_2$, and $\tau_1-\tau_2$, where the times are referenced with respect to the position of the third pulse. The pulse spacing $\tau_1$ refers to the first and second pulse, and $\tau_2$ refers to the second and third pulse. This is true for both two-level system echoes (i.e. spin echoes), as well as the classical anharmonic echoes that we are treating here. Figure \ref{3pulse} shows a three-pulse echo sequence in jammed systems. The signal is the average of 10,000 systems composed of 1000 particles each, and is identical to Figure \ref{setup}a, with the addition of a third pulse at a later time. If the third pulse is placed prior to the first echo, then $\tau_1-\tau_2$ is positive, and all four echoes occur after the three pulses.

\begin{figure}
\begin{center}
\includegraphics[width=.48 \textwidth]{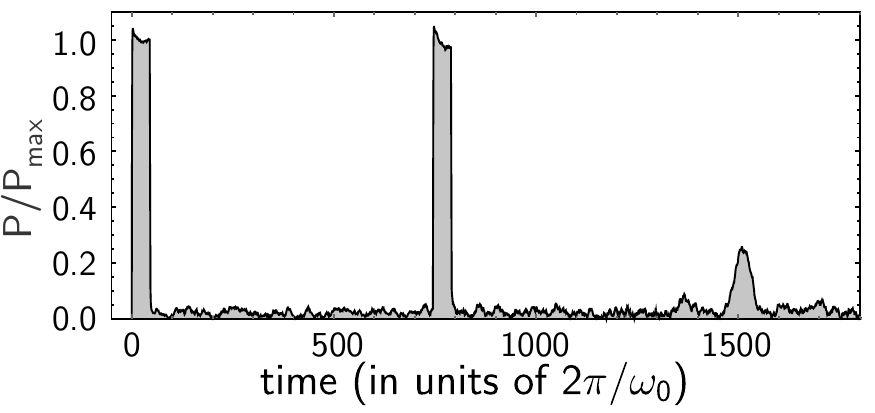}
\caption[]{Echoes in systems with Lennard-Jones interactions. The signal is the average of 10,000 independent systems, each composed of 1000 particles. The pulse width is $t_p$ = 45 cycles.} 
\label{LJechoes}
\end{center} 
\end{figure}

Although jammed systems of frictionless spheres provide the simplest example of model glasses, we have also studied echoes with more realistic two-particle potentials. Figure \ref{LJechoes} shows the average response of 10,000 systems, each composed of 1000 particles with Lennard-Jones interactions. Specifically, we use a Kob-Andersen binary system, as described in section \ref{NumericalModel}. The density of states of these systems looks somewhat different from jammed systems \cite{Xu2007}. The excitation frequency was chosen to be approximately 5\% of the maximum frequency in the system, and consistent with the longest-wavelength plane wave which could fit inside the simulation boundaries (Figure \ref{setup}b). The echo looks nearly identical to those in Figure \ref{setup}a and \ref{echovstau}a, where the particles interact via Hertzian potentials. 

\begin{figure}
\begin{center}
\includegraphics[width=.48 \textwidth]{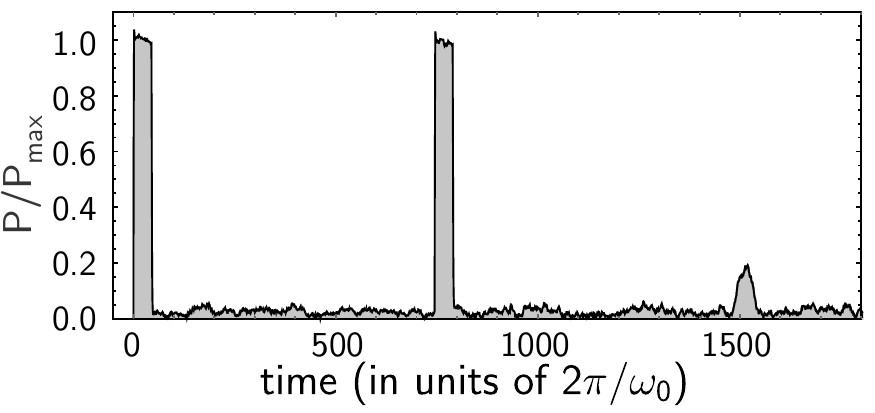}
\caption[]{Amplitude versus time showing an echo in systems of 8000 particles. The signal is the average of 5000 independent systems. The pulse frequency is half of that used in 1000-particle systems. The pulse width is $t_p$ = 45 cycles.} 
\label{8000particles}
\end{center} 
\end{figure}

All of the data that we have reported here so far has been taken on systems with 1000 particles, then averaged over many configurations. This is partially due to the fact that the calculation and characterization of the dynamical matrix and vibrational modes is straightforward. We can also average over fewer configurations of systems with a larger number of particles and obtain similar results. Figure \ref{8000particles} shows an echo signal resulting from averaging the response of 5000 systems, each composed of 8000 particles. Each cubic system of particles is twice as long on one side as a 1000 particle system, so the frequency of excitation was smaller by a factor of two, and the wavelength was longer by a factor of two (cf. Figure \ref{setup}b). 

However, for systems with more than 8000 particles, the computational requirements to observe an echo become expensive. This is mostly due to the fact that our simulations require simulating many thousands of cycles of low-frequency oscillations. However, we can estimate the conditions necessary to observe an echo in only one system, rather than averaging over many systems. In order to observe an echo clearly, there must be a sufficient number of excited modes, $N_e$, which will average to zero in regions between the pulses and the echo. The number of excited modes is proportional to the number of systems, $N_s$, the density of vibrational states at the excitation frequency, $\mathcal{D}(\omega)$, and inversely proportional to the pulse width, $t_p$:
\begin{align}
N_e\propto\frac{N_s\mathcal{D}(\omega)}{t_p}.
\label{Nefirst}
\end{align}
For simplicity, let us assume that the density of states obey's a Debye-like behavior, so that $\mathcal{D}(\omega)\propto N_p \omega^2$, where $N_p$ is the number of particles in the system. The pulse width used in our simulations is $t_p=45$ cycles $=45\times2\pi/\omega$. Then equation \ref{Nefirst} reduces to
\begin{align}
\label{Nesecond}
N_e\propto N_s N_p \left(\frac{\omega}{\omega_D}\right)^3. 
\end{align}

The frequency in equation \ref{Nesecond} has been normalized by the Debye frequency in order to easily compare to experiments. We are interested in comparing our simulations with experiments from low temperature glasses where possible. For that reason we chose $t_p=45$ cycles,  and $\omega/\omega_D\approx 0.0002$, which are typical values used in the original experiments which observed phonon echoes in glasses at low temperatures \cite{Graebner1979}. Using these values, $N_e\sim640$. We would need the same number of excited modes in a single system to see the echo. If we assume we have one system ($N_s=1$), then we would need $N_p\approx 8\times10^{13}$ particles to observe an echo at such low frequencies. We have also assumed that the density of states is quadratic in frequency. At temperatures below $T=100$ K, the density of states in glasses is known to decrease faster that $\omega^2$ \cite{Buchenau1986,Galperin1988,Galperin1989}. This would only strengthen the dependence of $N_e$ on $\omega$, and necessitate even larger systems in order to observe an echo, thus our estimate constitutes a lower bound on the system size.

%, which leads to more coupling between the vibrational modes, which can destroy the echo since the details depend on uncoupled, anharmonic oscillators (equation \ref{fastslow}). 

We would of course like to observe an echo in a single system, but this is computationally unfeasible. Not only does it require very large systems, but it also requires that the anharmonic oscillators do not couple the energy in between them strongly. In order to minimize this latter constraint, we suggest that we only apply frequencies in the region of the quasi-localized modes. 

The amount of coupling depends on the frequency difference between two modes: $\omega_{1}-\omega_{2}$. As we increase the number of particles, the density of states also increases, so the frequency spacing between the modes decreases. This is unavoidable. Also, the amount of coupling depends on the spatial overlap between the modes. Plane waves are extended modes that will inevitably share particle vibrations. However, two localized or quasi-localized modes, if sufficiently far away from each other, will have very little coupling, regardless of the frequency. 

Thus, given a fixed excitation frequency $\omega$, as $N_p\rightarrow\infty$, the energy flow between modes will eventually destroy the coherence of the echo. This can only be remedied if most of the excited modes are quasi-localized so that they can still behave as independent anharmonic oscillators. We suspect that echoes could be observed in a single, very large system provided that the density of plane waves is much smaller than the density of quasi-localized modes. 

%We believe this condition is true at the temperature ranges and frequencies used in echo experiments with molecular glasses ($T\ll1$ K).  

\section{Conclusions}

These results illustrate how the anharmonic vibrational modes in a jammed system of particles can give rise to phonon echoes, similar to those measured in glasses at low temperatures. The mechanism of echo generation is distinctly different from echoes produced by quantum-mechanical two-level systems \cite{Fossheim1982}. In our simulations, echoes are produced by a purely classical mechanism caused by the frequency shift of the anharmonic vibrational modes. This shift acts as a slowly-varying phase which evolves in the time between the pulses, resulting in a non-zero average of the ensemble.

%These results illustrate how the anharmonic vibrational modes in a jammed system of particles can give rise to phonon echoes, similar to those measured in glasses at low temperatures. The mechanism of echo generation is distinctly different from echoes produced by two-level systems \cite{Fossheim1982}. In our simulations, echoes are produced by the frequency shift of the anharmonic vibrational modes. This shift acts as a slowly-varying phase which evolves in the time between the pulses, resulting in a non-zero average of the ensemble. 

The anharmonicity of the vibrational modes can be studied at $T = 0$.  At low frequencies, model glasses based on jammed sphere packings have quasi-localized modes that contribute to the density of states \cite{Xu2010}.  It has been argued that jammed systems are marginally stable and inherently close to an instability where the structure will rearrange \cite{Wyart2005a,Wyart2012,Kallus2014}. The contribution of such incipient instabilities to the density of states is currently being evaluated \cite{Xu2016}. The anharmonicity that is important for echoes is generated by the frequency shift of a mode with increasing amplitude; this is due to expansion non-linearity \cite{Goodrich2014,Goodrich2014a}, and is measurable at small amplitudes. If the amplitude becomes too large, then there will be energy transfer between modes and the modes will lose coherence.  At even higher amplitude in a system with only finite-ranged interactions, the contacts can break and reform \cite{Schreck2011}.  This would destroy echo formation. 

%The anharmonicity of the vibrational modes can be understood from the $T=0$ vibrational spectrum in different model glasses, including jammed, frictionless spheres \cite{Goodrich2014}. The amorphous nature of glassy materials leads to an excess in the density of states at low frequencies, an excess which includes many quasi-localized modes. The quasi-localized modes provide a contribution to the density of states arising from incipient rearrangements to the vibrational spectrum \cite{Xu2016}. The anharmonicity of the modes has two different components. The anharmonicity that is important for echoes is generated by the frequency shift with mode amplitude, and is measurable at small amplitudes. At higher amplitudes the modes will couple to each other, possibly due to contact breaking and rearrangement \cite{Schreck2011}, which is a another component of anharmonicity which acts to thermalize the system.  

\begin{figure}
\begin{center}
\includegraphics[width=.48 \textwidth]{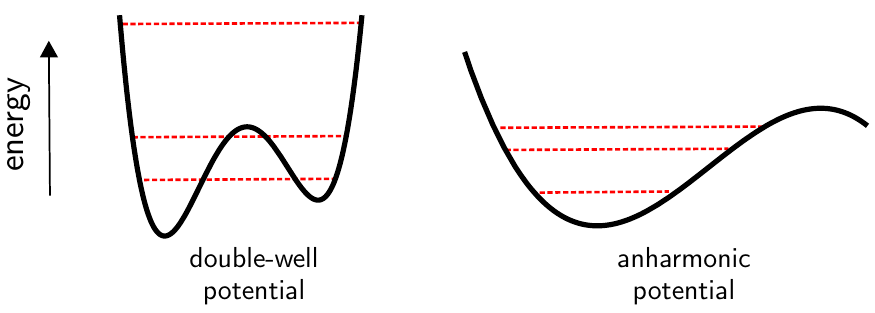}
\caption[]{(Color online) Quantum mechanical picture of energy levels corresponding to the low-frequency excitations in glasses. The double-well is the traditional view of the origin of two-level systems, where the lowest energy level is split in two by the presence of the barrier. Alternatively, a wide, anharmonic potential can produce low-frequency modes, non-uniform level spacing, and nonlinear acoustic phenomena such as echoes. } 
\label{potentials}
\end{center} 
\end{figure}

The echoes observed in our simulations have many features which are consistent with parametric, two-level system echoes, such as the three-pulse echo sequence (Fig.\ \ref{3pulse}). However, many features are quite different. First, anharmonic echoes do not have a simple, intuitive condition for maximizing the echo signal, such as a $\pi/2$ pulse followed by a $\pi$ pulse. In fact, Figure \ref{analytic} shows that the maximum echo amplitude is a complicated function of pulse spacing and amplitude. However, here we note that for small amplitudes, equation \ref{absechoamp} reduces to:
\begin{align}
|G_{2\tau}|=&A_1\alpha^2\Omega\tau=A_1A_2^2\xi\omega\tau,
\end{align}
which is in agreement with previous authors \cite{Herrmann1967}, and has the same dependence on $A_1$ and $A_2$ as the small-amplitude result for spin echoes \cite{Graebner1979}.

Graebner and Golding \cite{Graebner1979} measured this small-amplitude dependence in silica glass, and also showed that the maximum echo intensity does not precisely occur when $A_1=A_2/2$, among other discrepancies with a model of echoes based on two-level systems. In addition, at small amplitudes, Graebner and Golding observed a small increase in echo amplitude with pulse spacing. This feature is characteristic to anharmonic echoes and is seen in our simulations (Fig.\ \ref{echovstau}b). However, the experiments probe $\approx$ 1000 times lower frequencies than we can access in the simulations, and also involve additive reflections of pulses, so a more quantitative comparison is complicated. Qualitatively, we note that the appearance of multiple echoes after two excitation pulses in the echo experiments \cite{Graebner1979} is a natural and unique feature in anharmonic echoes, and does not depend on the details of the system.

At very low temperatures, a quantum mechanical picture of the dynamics is certainly necessary. The traditional explanation for the excess excitations in glasses at very low temperatures relies on two-level tunneling systems created by the splitting of the ground state energy in a double-well potential (Fig.\ \ref{potentials}). The distribution of the energy barriers are assumed to be broad, and the states are spatially localized. We offer an alternative picture based on localized, anharmonic vibrational modes that can be understood both classically and quantum mechanically. A wide and shallow anharmonic potential (Fig.\ \ref{potentials}), characterized by low-frequency, nearly-unstable modes, will have energy levels which are not equally spaced. These modes arise naturally due to the amorphous nature of the solid, and do not depend on specific particle interactions. 

Our results have focused solely on the origin of phonon echoes in glasses, and have not addressed many other well-known nonlinear acoustic properties in glasses, such as saturation of attenuation and hole-burning. However, the existence of these phenomena in glasses may not be restricted to models that require two-level systems. Past theoretical results suggest that some universal thermodynamic properties in glasses need only modest assumptions about the nature of the low-temperature modes \cite{Yu1988}. More recent results show that universal features of acoustic attenuation in glasses can be explained by generic, elastically-coupled resonant modes, and that the details and origins of the resonant modes are less important \cite{Vural2011}. 

Finally, we note that the simulations presented here were performed on small systems. The detailed properties of the anharmonic modes at very low-frequencies have yet to be investigated due to computational limitations on system size. One benefit of using jammed spheres with finite-ranged repulsions as a model glass is that there are two limiting regimes to investigate \cite{Goodrich2014}. Although the limit $N\rightarrow\infty$ is inaccessible, we can take the limit $\Delta\phi\rightarrow 0$, bringing the system on the verge of instability. In this regime it is well known that jammed systems develop an enormous increase in the density of states. The fate of these excess modes as the the system is compressed above $\Delta\phi=0$ provides a starting point for our understanding of the low-frequency, anharmonic modes. 

\section{Acknowledgements}

We are grateful to Brent Busby for assistance with computational servers.  We thank Efi Efrati, Carl Goodrich, Dustin Kleckner, Andrea Liu, and Ning Xu for important discussions. We acknowledge support from NSF MRSEC DMR-1420709, NSF DMR-1455086 (J.C.B.), and the US Department of Energy, Office of Basic Energy Sciences, Division of Materials Sciences and Engineering, Award No. DE-FG02-03ER46088 (S.R.N.).

\bibliography{echoes}

\end{document}